\documentclass[useAMS,usenatbib]{mn2e}

\usepackage{mathptmx}
\usepackage{multirow}
\usepackage{psfrag}
\usepackage{pspicture}
\usepackage{graphicx}
\usepackage{amsmath}
\usepackage{color}
\usepackage{ amssymb }

\title[The widest-frequency relic spectra]{The widest-frequency radio relic spectra: observations from 150 MHz to 30 GHz}
\author[A. Stroe et al.]{Andra Stroe$^{1}$\thanks{E-mail: astroe@eso.org; Present address: European Southern Observatory, Karl-Schwarzschild-Str. 2, 85748, Garching, Germany}, 
Timothy Shimwell$^1$, Clare Rumsey$^{2}$, Reinout van Weeren$^{3}$, Maja Kierdorf$^{4}$,
\newauthor Julius Donnert$^{1}$, Thomas W. Jones$^5$, Huub J. A. R\"ottgering$^{1}$,  Matthias Hoeft$^6$, 
\newauthor Carmen Rodr{\'i}guez-Gonz{\'a}lvez$^7$, Jeremy J. Harwood$^{8}$, Richard D. E. Saunders$^{3,9}$\\
$^{1}$Leiden Observatory, Leiden University, P.O.\ Box 9513, NL-2300 RA Leiden, The Netherlands\\
$^{2}$Astrophysics Group, Cavendish Laboratory, JJ Thomson Avenue, Cambridge, CB3 0HE\\
$^{3}$Harvard Smithsonian Center for Astrophysics (CfA - SAO), 60 Garden Street Cambridge, MA 02138, US\\
$^4$Max-Planck-Institut f{\"u}r Radioastronomie, Auf dem H{\"u}gel 69, D-53121 Bonn, Germany\\
$^{5}$School of Physics and Astronomy, University of Minnesota, Minneapolis, MN 55455, USA\\
$^{6}$Th\"uringer Landessternwarte (TLS), Sternwarte 5, D-07778 Tautenburg, Germany\\
$^{7}$U.S. Planck Data Center, MS220-6, Pasadena, CA 91125, USA\\
$^{8}$ASTRON, Postbus 2, 7990 AA Dwingeloo, The Netherlands\\
$^{9}$Kavli Institute for Cosmology Cambridge, Madingley Road, Cambridge CB3 0HA, UK\\
\vspace{-15pt}
}
\begin{document}
\maketitle
\begin{abstract}
Radio relics are patches of diffuse synchrotron radio emission that trace shock waves. Relics are thought to form when intra-cluster medium electrons are accelerated by cluster merger induced shock waves through the diffusive shock acceleration mechanism. In this paper, we present observations spanning $150$ MHz to $30$ GHz of the `Sausage' and `Toothbrush' relics from the Giant Metrewave and Westerbork telescopes, the Karl G. Jansky Very Large Array, the Effelsberg telescope, the Arcminute Microkelvin Imager and Combined Array for Research in Millimeter-wave Astronomy. We detect both relics at $30$ GHz, where the previous highest frequency detection was at $16$ GHz. The integrated radio spectra of both sources clearly steepen above $2$ GHz, at the $\gtrsim6\sigma$ significance level, supports the spectral steepening previously found in the `Sausage' and the Abell 2256 relic. Our results challenge the widely adopted simple formation mechanism of radio relics and suggest more complicated models have to be developed that, for example, involve re-acceleration of aged seed electrons.
\end{abstract}
\begin{keywords}
acceleration of particles, radiation mechanisms: non-thermal, shock waves, galaxies: clusters: individual: CIZA J2242.8+5301, 1RXS J0603.3+4214, radio continuum: general
\vspace{-15pt}
\end{keywords}

\section{Introduction}\label{sec:intro}
Radio relics are polarized areas of diffuse, low-brightness radio emission often with arc-like morphologies and typical $>1$ Mpc size \citep{2012A&ARv..20...54F,2014IJMPD..2330007B}. Relics are placed exclusively at the outskirts of massive, post-core passage merging galaxy clusters and preferentially oriented perpendicularly to the merger axis \citep{2012A&ARv..20...54F}. Whilst there are spectral index variations within individual relics \citep[usually across the relic width, e.g.][]{2007A&A...467..943O, 2010Sci...330..347V, 2012A&A...546A.124V, 2012MNRAS.426...40B, 2013A&A...555A.110S} the integrated radio relic spectra below $5$ GHz are well described by a single power law with $\alpha<-1$, where radio spectral index $\alpha$ is described as the flux density $F$ as function of the frequency $\nu$: $F \sim \nu^{\alpha}$ \citep{2012A&ARv..20...54F}. The synchrotron nature of these sources, coupled with the strong polarization, indicates a significant ordering in the magnetic field structure: the magnetic field vectors are aligned with shock structure \citep[e.g.]{2010Sci...330..347V, 2012A&A...546A.124V}. The observational results regarding the morphology, spectrum and polarisation of radio relics led to a favoured interpretation. In this scenario, relics are produced by synchrotron-emitting cosmic ray electrons accelerated by shocks through the diffusive shock acceleration mechanism \citep[DSA;][]{1998A&A...332..395E}. Cluster merger events produce such weak shock waves in the ICM (Mach number $M<5$), where part of the gravitational energy released during the merger event is dissipated \citep[e.g.][]{2006MNRAS.367..113P}.

However, more recently, evidence has been found in tension with the simple DSA picture, as it was proposed by \citet{1998A&A...332..395E}. Simulations have found that producing substantial radio flux densities at low Mach number shocks with electrons accelerated out of the local thermal population would require extraordinarily large particle injection efficiencies \citep[e.g.][]{2007ApJ...669..729K}. Furthermore, robustly detected X-ray shocks were found to have no radio counterpart or be associated with a radio shock at an offset position \citep[e.g.][]{2011MNRAS.417L...1R, 2013MNRAS.433..812O, 2014MNRAS.443.2463O,2014MNRAS.440.2901S}, while in cases with a joint shock detection, a discrepancy between the X-ray and the radio Mach number measurement was found \citep{2012A&A...546A.124V, 2013MNRAS.429.2617O}. However, the initially measured discrepancy between the two Mach number estimates for the `Sausage' relic in cluster CIZA J2242.8+5301, has since been partially attributed to resolution effects affecting the radio measurement \citep{2014MNRAS.445.1213S}. 

High frequency flux measurements are useful to help resolving these issues. Very recently, high frequency radio measurements of relics have been published for the first time at $16$ GHz with the Arcminute Microkerlvin Imager \citep[AMI, ][]{2014MNRAS.441L..41S} and the Effelsberg telescope at $10$ GHz \citep{2015A&A...575A..45T}. Both studies find evidence for steepening of the radio spectrum at frequencies higher than $2$ GHz. \citet{2015MNRAS.447.2497E} proposed that the Sunyeav-Zeldovich effect \citep[SZE;][]{1972CoASP...4..173S}, the upscattering of cosmic microwave background photons by the intra-cluster medium electrons, could be responsible for the perceived steepening. Relics would sit in a negative `bowl', which would result in a measured relic flux density smaller than the real value. \citet{2014MNRAS.441L..41S} propose that the cause for the steepening lies in the physics of the relic. A non-power-law injection spectrum could lead to a curved integrated index. Furthermore, if the magnetic field is ordered and boosted at the shock location, but turbulent in the downstream area, the integrated spectrum could steepen even more at high frequencies. \citet{2015arXiv150504256K}  use time-dependent DSA simulations of a spherical shock impinging on a magnetised cloud of pre-accelerated relativistic electrons to obtain relic spectra which gradually steepen over the $(0.1 - 10)$ GHz range. However, the steepening is insufficient to fully explain the measurements by \citet{2014MNRAS.441L..41S}. Testing theoretical models has been challenging due to a dearth of measurements above $2.5$ GHz. For example, \citet{2015A&A...575A..45T} have integrated flux density measurements of the Abell 2256 relic only at $5$ and $10$ GHz, while \citet{2014MNRAS.441L..41S} have one measurement for the `Sausage' relic, but at a much larger lever arm ($16$ GHz). This means the actual shape of the high-frequency spectrum is not well constrained.

The clusters CIZA J2242.8+5301 \citep[`Sausage';][]{2007ApJ...662..224K, 2010Sci...330..347V} and 1RXS J0603.3+4214 \citep[`Toothbrush';][]{2012A&A...546A.124V} are ideal targets as both clusters host a bright relic ($F>0.15$ Jy at $1.4$ GHz). Both relics are located towards the northern periphery of the clusters, extending over more than $1.5$ Mpc in length, but less than $200$ kpc in width. The rich data below $2.5$ GHz on the `Sausage' and `Toothbrush' relics were of sufficient quality to enable \citet{2012A&A...546A.124V}, \citet{2013A&A...555A.110S} and \citet{2014MNRAS.445.1213S} to find gradients of increasing spectral index and spectral curvature from the northern towards the southern edge of the relics. \citet{2014MNRAS.445.1213S} fitted spectral ageing models to the `Sausage' data to interpret the trends as increasing electron age across the relic.

We have performed new high-frequency observations of the two relics with the Karl G. Jansky Very Large Array (VLA) in the $2-4$ GHz range, with the Effelsberg telescope at $5$ and $8$ GHz, with AMI at $16$ GHz and with the Combined Array for Research in Millimeter-wave Astronomy (CARMA) at $30$ GHz. We combine the new observations with the radio data already available from the Giant Metrewave Radio Telescope (GMRT), the Westerbork Synthesis Radio Telescope (WSRT) and AMI \citep{2012A&A...546A.124V, 2013A&A...555A.110S, 2014MNRAS.441L..41S} to produce a well-sampled spectrum for the main radio relics over a frequency range spanning $2.3$ dex. 

Assuming a flat, $\Lambda$CDM cosmology with $H_0 = 70$ km s$^{-1}$, matter density $\Omega_\mathrm{m} = 0.27$, dark energy density $\Omega_\mathrm{\Lambda}=0.73$, at the redshift of the two clusters, $z\sim0.2$, $1$ arcmin corresponds to a scale of $\sim0.2$~Mpc. All images are in the J2000 coordinate system. 

\section{Observations and data reduction}
\label{sec:obs-reduction}
For our analysis we combine existing observations with new data for the `Sausage' and `Toothbrush' relics. The frequencies of the observations can be found in Table~\ref{tab:freqs}. Standard calibration was applied to all the data sets, including flagging, bandpass and gain calibration. The details of the reduction of the existing GMRT and WSRT datasets can be found in \citet{2012A&A...546A.124V} and \citep{2013A&A...555A.110S}. 

\begin{table}
\begin{center}
\caption{Summary of the frequencies at which observations have been made for the `Sausage' and `Toothbrush' radio relics. In both cases, the observations span from $150$ MHz to $30$ GHz.}
\vspace{-10pt}
\begin{tabular}{l c c}
\hline
\hline
Source & Telescope & Frequencies \\
\hline
\multirow{6}{*}{`Sausage'} & GMRT & $150$, $325$, $610$ MHz \\
\multirow{6}{*}{15 frequencies} & WSRT & $1230$, $1380$, $1710$, $2270$ MHz \\
& VLA & $2250$, $2750$, $3250$, $3750$ GHz \\
& Effelsberg & $4.85$, $8.35$ GHz \\
& AMI & $15.85$ GHz \\
& CARMA & $30$ GHz \\ \hline
\multirow{5}{*}{`Toothbrush'} & GMRT & $150$, $240$, $324$, $608$ MHz \\ 
\multirow{5}{*}{12 frequencies} & WSRT & $1230$, $1380$, $1710$, $2270$ MHz \\
& Effelsberg & $4.85$, $8.35$ GHz \\
& AMI & $15.85$ GHz \\
& CARMA & $30$ GHz \\
\hline
\end{tabular}
\label{tab:freqs}
\vspace{-10pt}
\end{center}
\end{table}

\subsection{Jansky VLA data}
Observations of the `Sausage' cluster in S band were performed in D array configuration with the VLA on 27 January 2013. Data were recorded in full polarisation, in 16 spectral windows of 128 MHz each, spanning a bandwidth of $2$ GHz between $2$ and $4$ GHz. Each spectral window was further subdivided into 64 channels. The $1.5$ h on source-time were equally divided over three separate pointings to cover the cluster region. Two primary calibrators were observed, 3\,C138 and 3\,C147. J2202+4216 was used as a secondary calibrator. The pointing centres for the cluster were: RA=22:43:19, DEC=+53:05:28; RA=22:42:36, DEC=+53.07:29 and RA=22:42:41, DEC=+52:58:09. The data were calibrated with the Common Astronomy Software Applications package\footnote{http://casa.nrao.edu/} \citep[\textsc{casa} version 4.3][]{2007ASPC..376..127M}. The data were first corrected for the antenna offset positions and elevation dependent gain curves. The data were also Hanning smoothed and strong radio frequency interference (RFI) was removed with the `tfcrop' option of the task flagdata(). The flux scale for the primary calibrators was set using the calibrator model image provided by casapy and taking the \citet{2013ApJS..204...19P} scale. We then determined an initial bandpass correction using the primary calibrator sources. This is done to remove the strong bandpass rolloff at the edges of the spectral windows which hinders the detection of RFI. RFI was then removed with \textsc{AOFlagger} \citep{2010MNRAS.405..155O}. After RFI removal, we determined initial gain corrections using the central 10 channels of the spectral windows. These corrections were pre-applied before finding the delay solutions and subsequent final bandpass solutions. By pre-applying the gain solutions we remove any temporal variations in the gains which would otherwise effect the bandpass and delay terms. The gain solutions were then re-determined for all calibrator sources but now pre-applying the bandpass and delay solutions. Next we bootstrapped the primary flux scale to the secondary calibrator. The final calibration solutions were applied to the target fields. 

\begin{figure}
\begin{center}
\includegraphics[trim=0cm 0cm 0cm 0cm, width=0.495\textwidth]{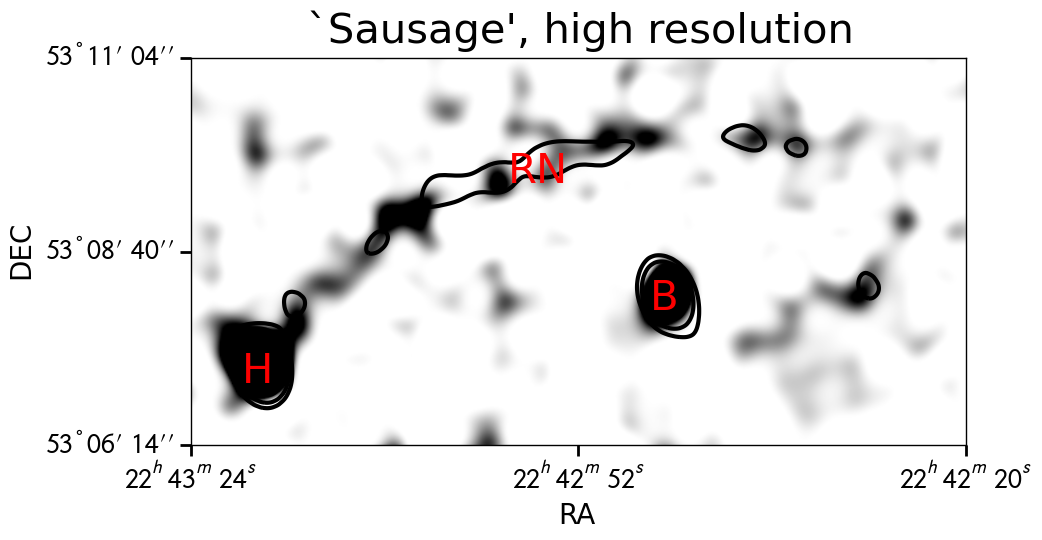}\\ \vspace{5pt}
\includegraphics[trim=0cm 0cm 0cm 0cm, width=0.495\textwidth]{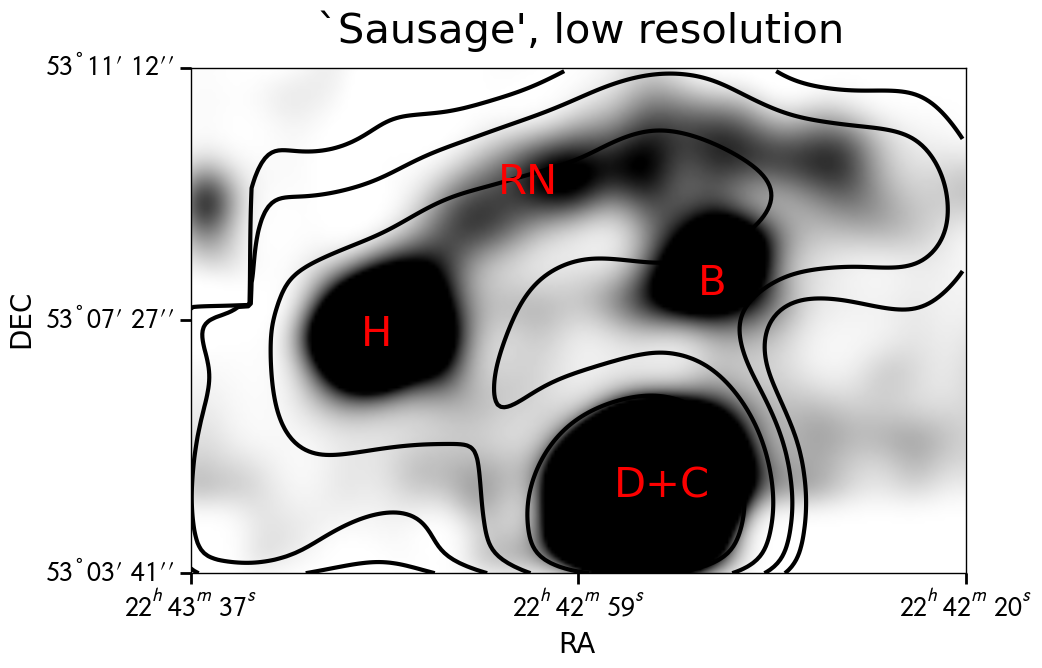}
\end{center}
\vspace{-10pt}
\caption{Low resolution total power and uniform-weighted high resolution images of the `Sausage' cluster. The top panel shows the CARMA image ($\sim30$ arcsec resolution) in grayscale and the AMI-LA ($\sim40$ arcsec beam) image in contours drawn at ${[4,8,16,32]} \times \sigma_{\mathrm{RMS}}$. The bottom panel shows the Effelsberg $8.35$ GHz ($90$ arcsec resolution) image in grayscale and the $4.85$ GHz ($159$ arcsec resolution) in contours at ${[4,8,16,32]} \times \sigma_{\mathrm{RMS}}$. We label the source as \citet{2013A&A...555A.110S}. RN refers to the relic, which B, D, C and H and radio galaxies.}
\label{fig:mapS}
\vspace{-10pt}
\end{figure}

\begin{figure}
\begin{center}
\includegraphics[trim=0cm 0cm 0cm 0cm, width=0.495\textwidth]{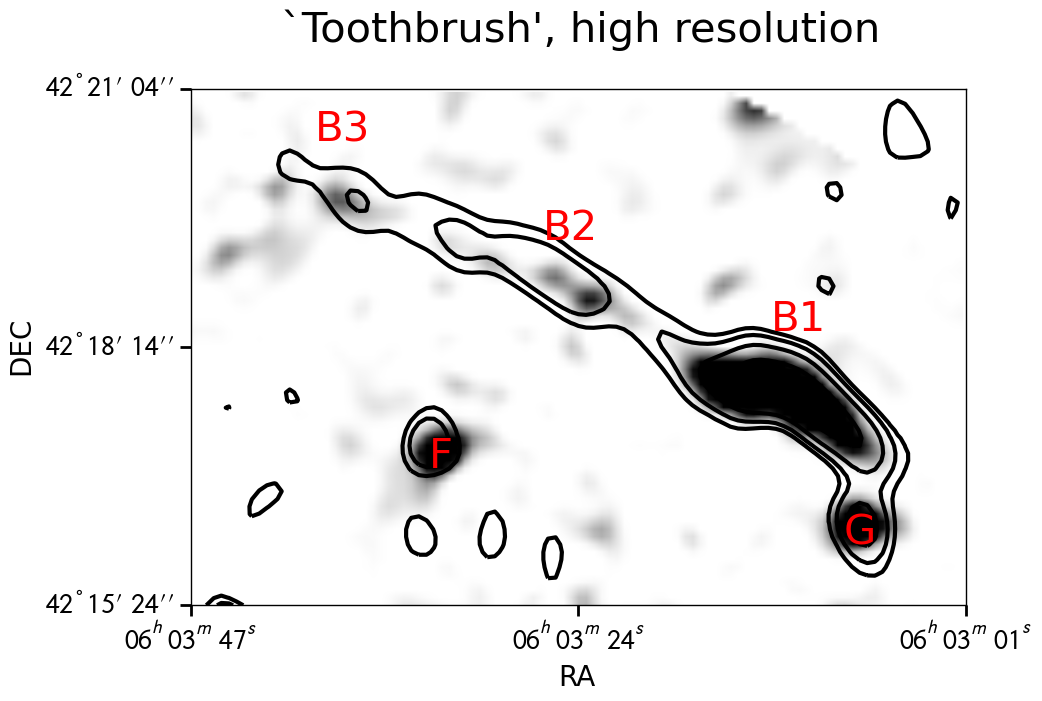}\\ \vspace{5pt}
\includegraphics[trim=0cm 0cm 0cm 0cm, width=0.495\textwidth]{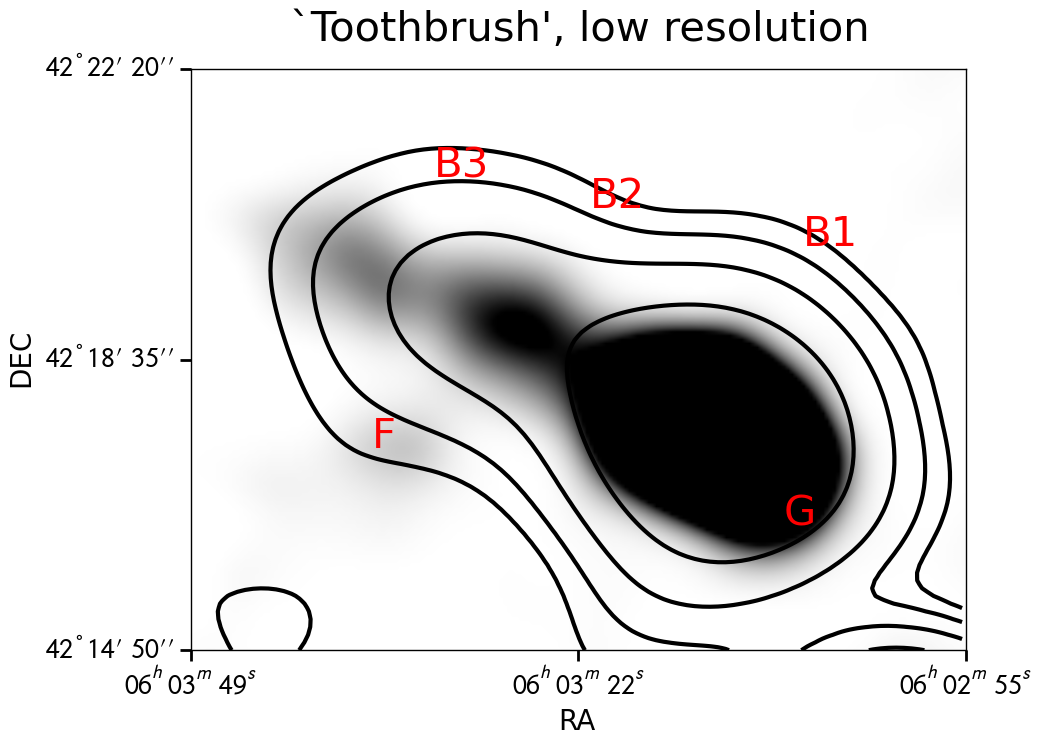}
\end{center}
\vspace{-10pt}
\caption{Same as Figure \ref{fig:mapS}, but for the `Toothbrush' relic. B1, B2 and B3are subsections of the relic, labelled according to \citet{2012A&A...546A.124V}. Radio galaxy F \citep[from][]{2012A&A...546A.124V} and G are also marked.}
\label{fig:mapT}
\vspace{-10pt}
\end{figure} 

\subsection{Effelsberg data}
In the following section we describe the Effelsberg observations, data reduction and flux density measurements in detail.

\subsubsection{Effelsberg observations}
The observations were performed with the Effelsberg 100-m Radio Telescope in October/November 2010, January 2011, October 2011 and August 2014 using the 3.6 cm (8.35 GHz with $1.1$ GHz bandwidth) single-horn and 6.3 cm ($4.85$ GHz with $0.5$ GHz bandwidth) dual-horn receiving systems installed at the secondary focus. The radio sources 3C\,48, 3C\,138, 3C\,147, 3C\,286 and 3C\,295 were observed as flux density and pointing calibrators, using the flux densities from Peng et al. (2000). 

Pointing checks were performed at regular intervals. A focus measurement was performed to account for temperature variations by moving the secondary mirror. We measured a calibration map for the flux density and polarization calibration. A well-known, bright, not variable and unresolved radio source with constant flux density was used as a calibrator.

The 8.35 GHz observations were made with a single-horn system, located in the secondary focus of the telescope. The desired area were scanned alternatively in RA and DEC directions, where the sub-scans were separated by 30 arcsec. Baselevels were subtracted automatically by linear interpolation between pixels at both sides of each sub-scan. In the end, we obtained 46 maps for `Sausage' and 40 for `Toothbrush'. The final map sizes for the `Sausage' and `Toothbrush' fields were $10'\times14'$ and $10'\times15'$, respectively.

The 4.85 GHz observations were made with a double-horn system. As the horns are fixed in the secondary focus, scanning has to be done in azimuth to allow the restoration of the signals \citep{1979A&A....76...92E}. The maps need to be larger in azimuth by the beam separation of 8 arcmin, plus one additional beamwidth on each side to improve the baselevel determination. The sub-scans were separated by 60 arcsec and the baselevels were subtracted. We obtained 32 maps for `Sausage' and 12 for `Toothbrush' scanned in azimuthal direction. The final map sizes for the `Sausage' and `Toothbrush' fields at 4.85 GHz were $16'\times26'$ and $20'\times30'$, respectively.

\begin{table*}
\begin{center}
\caption{Imaging parameters and data used in combination for producing integrated spectra. For the interferometric spectrum made from comparable datasets, all the data were convolved to the image with the lowest resolution (imposed by the AMI-LA data). For the total-power measurements the data were left at their native resolution (best resolution obtained by the interferometric data with the robust weighting). Convolving the images to the lowest resolution of Effelsberg (the largest beam is 2.5 arcmin at 4.85 GHz) would result in heavy blending of diffuse flux with point sources.}
\vspace{-10pt}
\begin{tabular}{l c c c c c p{3.0cm} p{4cm}}
\hline
\hline
Source & Type & uv range & Weighting & Max scale & Resolution & Telescope & Frequencies \\
& & & k$\lambda$ & $'$ & $''$ & & GHz \\
\hline
\multirow{2}{*}{`Sausage'} & Total power & all & robust  & - & $5-147$ & GMRT, WSRT, Effelsberg & $0.15$, $0.325$, $0.61$, $1.2$, $4.85$, $8.35$ \\ 
& Interferometric & $>0.8$ & uniform & $\sim4$ & $40$ & GMRT, WSRT, VLA, AMI-LA, CARMA &  $0.15$, $0.325$, $0.61$, $1.2$, $1.4$, $1.7$, $2.25$, $2.3$, $2.75$, $3.25$, $3.75$, $15.85$, $30$ \\ \hline
\multirow{2}{*}{`Toothbrush'} & Total power & all  & robust & - & $5-147$ & GMRT, WSRT, Effelsberg & $0.15$, $0.24$, $0.325$, $0.61$, $1.2$, $4.85$, $8.35$ \\
& Interferometric & $>0.8$ & uniform & $\sim4$ & $40$ & GMRT, WSRT, AMI-LA, CARMA & $0.15$, $0.24$, $0.325$, $0.61$, $1.2$, $1.4$, $1.7$, $2.3$, $15.85$, $30$  \\ \hline
\hline
\end{tabular}
\label{tab:imaging}
\vspace{-10pt}
\end{center}
\end{table*}

\subsubsection{Effelsberg data reduction}

 The raw Effelsberg maps include artefacts which need to be removed. Every single map was checked and edited if there were disturbances which result from weather effects or other problems such as, RFI, scanning effects due to clouds and baselevel distortions due to sources near the edge of a map. The main part of data reduction was accomplished using the NOD2-based software package called `Ozmapax' \citep{1974A&AS...15..333H}. The software allows us to check individual maps for disturbances and edit them.

For the $8.35$ GHz maps it was necessary to check every map for disturbances (RFI, scanning effects or baselevel distortions). In case of scanning effects or RFI, the entire affected scans must be set to dummy values. If a source is located near the edge, the baselevel of the entire scan can be incorrect. To make a baselevel modification one must define a new baselevel in a region further away from the source. All this has to be done separately for Stokes I, Q and U. After editing and separating all maps, we combined all reduced maps to one final map in Stokes I, Q and U using the basket-weaving method \citep{1988A&A...190..353E}.

For the $4.85$ GHz total power (Stokes I) data weather effects were removed by creating the difference of the maps from the two different horns. The result was a rectangular map in which the source appears twice: once with positive values and once with negative ones. Most weather effects were eliminated in the difference map. Nevertheless, it was necessary to correct all difference maps for residual disturbances. After editing all Stokes Q and U (see above) and all difference maps (Stokes I) we combined them to improve the  signal to noise (S/N). 

Further reduction steps were done by the NRAO \textsc{AIPS} \footnote{http://www.aips.nrao.edu} (Astronomical Image Processing System) software package. All Stokes maps are convolved to $90$ arcsec at $8.35$ GHz and $159$ arcsec at $4.85$ GHz to increase S/N ratios. These maps will be fully presented and discussed in an upcoming paper (Kierdorf et al. in preparation).

\subsection{AMI data}
The AMI telescope \citep{2008MNRAS.391.1545Z} consists of two interferometers: the Large Array (AMI-LA) with baselines of 18-110 m has a resolution of approximately $30$ arcsec; the shorter Small Array (AMI-SA) baselines of 5-20 m give a resolution of about $3$ arcmin. The `Sausage' AMI-LA data and its reduction are summarised in \citet{2014MNRAS.441L..41S}. Initial AMI-LA observations for the `Toothbrush' cluster were taken at $\sim16$ GHz with 61 pointings to cover the entire cluster area. The relic was subsequently observed in a 4-point mosaic on 7 November 2013. We follow the same reduction steps for the `Toothbrush' as described in \citet{2014MNRAS.441L..41S} for the `Sausage' cluster.

\subsection{CARMA data}

The `Sausage' and `Toothbrush' relics were observed with the CARMA 6.1\,m and 10.4\,m dishes in the compact E-configuration  during 2014 July 9 to 2014 August 3 (project c1223). The data were recorded with eight 500\,MHz frequency bands placed between 27.4 and 32.7\,GHz where known `birdies' (radio chirps caused by harmonics) in the 1cm system were avoided . To obtain close to uniform sensitivity across the large relics we observed using  11-point mosaics where pointings were separated by 1.4 arcmin which is sufficiently close to Nyquist sample the pointings at the highest observed frequency for the largest antenna. A total of 14\,hrs and 21\,hrs hours of on-source data were recorded for the `Toothbrush' and `Sausage' clusters, respectively, and observations were performed over a range of elevations to provide good uv-coverage. During an observation each pointing in the mosaic was observed for 1 minute before a calibrator was observed for three minutes, this cycle was repeated until the end of the observation. We used BL Lac as the interleaved gain calibrator for the `Sausage' relic and 055+398 for the `Toothbrush' cluster. During each observation a passband calibrator was also observed for 10 minutes, we used 0510+180 and 0927+390 for the `Toothbrush' observations and MWC349 for the `Sausage' observations. The calibration was done with the \textsc{miriad} package (\citealt{sault_1995}) following the procedure described in \cite{shimwell_2013}. 

\section{Imaging and flux density measurements}

The goal for our analysis is to determine the integrated spectrum of the two radio relics. We make two sets of measurements: one using interferometric images made from comparable data using the same inner uv cut and same resolution, the second using the interferometric measurement with the best uv coverage going to low uv spacings in an attempt to combine the flux density measurement with total power flux densities as measured by Effelsberg.

\subsection{Interferometric measurements}
The integrated flux density at each frequency is measured from the map. Each data set is imaged with uniform weighting, sampling the same spatial scales of the sky.
Such images can be produced if we have the same uv coverage in each interferometric observation. In practice, we approximate this by applying a uv range and a uniform weighting to visibilities when imaging. We image the data using the CLEAN algorithm with the same resolution, the same pixel ($1$ arcsec per pixel) and image size and correct all images for the effects of the primary beam attenuation.

\subsubsection{Radio images}\label{sec:res:images}

We produced interferometric images using the CLEAN algorithm and refined them through the process of self-calibration. In the case of the VLA data, we decided to split the data into four $0.5$ GHz wide chunks to allow flux density measurements at central frequencies of 2.25, 2.75, 3.25, and 3.75 GHz with sufficient S/N and without being hindered too much by spectral variations within a 0.5 GHz bandwidth. For these four VLA images, the three pointings were jointly convolved and imaged, correcting for the VLA primary beam scaled for the central frequency and using the `mosaic' option of the clean() task in \textsc{casa}. To facilitate imaging from the heterogeneous CARMA array we used the \textsc{casa} package with the default CARMA primary beams. As for the VLA data, we used the `mosaic' mode in the imager to jointly deconvolve the entire `Sausage' and `Toothbrush' mosaics CLEANing to a depth of 0.15\,mJy/beam. To improve the image quality our data were self calibrated with two rounds of phase-only calibration, followed by two final rounds of amplitude and phase calibration. The amplitude solutions were normalized to 1.0 to prevent drifting of the flux scale. For imaging during the self-calibration we used Briggs weighing with a robust factor of 0.0 \citep{briggs_phd}. To aid the deconvolution we used CLEAN boxes. 

\begin{figure*}
\begin{center}
\includegraphics[trim=0cm 0cm 0cm 0cm, width=0.995\textwidth]{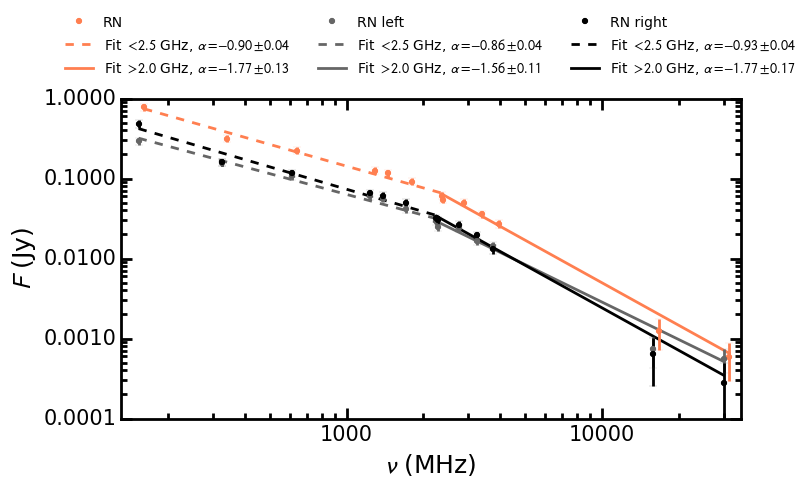}
\end{center}
\vspace{-10pt}
\caption{Integrated spectrum of the `Sausage' radio relic measured at $\sim30$ arcsec resolution from $150$ MHz to $30$ GHz.  There is clear evidence for spectral steeping beyond $2.5$ GHz. A single power law does not fit the data, while a broken power law provides a much better description. Note the results hold even when we split the source in two halves. The total relic flux values are slightly shifted to the right, for a better visibility of the values and errors.}
\label{fig:intspecS}
\vspace{-10pt}
\end{figure*}

\begin{figure*}
\begin{center}
\includegraphics[trim=0cm 0cm 0cm 0cm, width=0.995\textwidth]{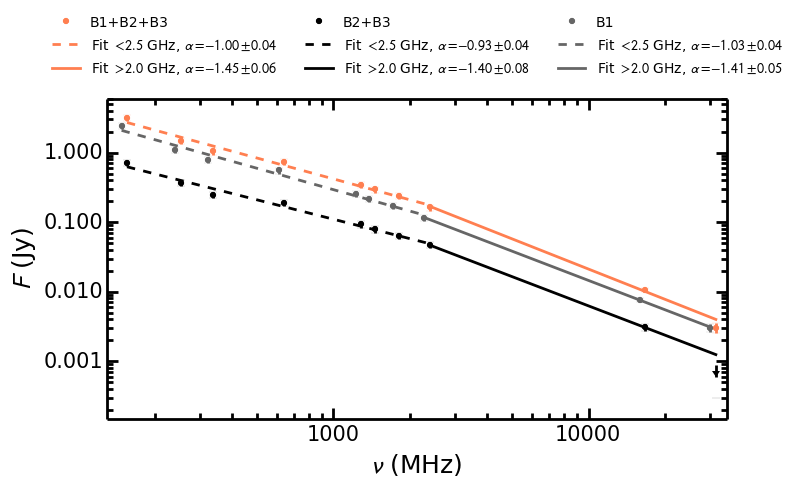}
\end{center}
\vspace{-10pt}
\caption{Same as Figure \ref{fig:intspecS}, but for the `Toothbrush' relic. Note that B2+B3 are not detected in the CARMA image, hence a $3\sigma$ upper limit is drawn.}
\label{fig:intspecT}
\vspace{-10pt}
\end{figure*}

We produce primary beam corrected images using the same uv ranges from our interferometric data, in order to study the diffuse emission on different scales. The details of the imaging parameters and the data used can be found in Table~\ref{tab:imaging}. We calculate the largest spatial scale fully recovered using the conversion: $0.6\lambda/b_\mathrm{min}$, $\lambda$ is the wavelength and $b_\mathrm{min}$ is the smallest baseline sampled in the uv plane\footnote{Note that formally an interferometer is a matched filter and it is sensitive to emission on scales of $\sim\lambda/ b_\mathrm{min}$. However, assuming a Gaussian brightness distribution the flux on scales of $\sim0.6\lambda/ b_\mathrm{min}$ can be fully recovered, while emission on scales up to $\sim\lambda/ b_\mathrm{min}$ is partially recovered. This estimation comes from Fourier transforming the Gaussian source brightness distribution and estimating on what angular scale the recovered power drops to $\sim50$ per cent.}. To probe spatial scales of up to $\sim2.6$ arcmin, we use only data at uv distances beyond $800\; \lambda$ and convolve the images to $\sim0.5$ arcmin resolution, as imposed by the AMI-LA uv coverage. We probe the uv space over a factor of 8 in baseline length, ranging from $800\,\lambda$ and at least up to $\sim6600\,\lambda$. Note that any flux at scales larger than this is mostly aligned with the shock (east-west direction). This means we are not resolving out any flux along the north-south direction (because there is no diffuse relic component in this direction). Therefore in the downstream direction we are capturing all the emission. For a given shock section we are capturing all the emission coming from the recently shock-accelerated electrons as well as the aged particles. Therefore the uv cut are not affecting the integrated flux density measurements, in the sense of biasing its shape because of preferential selection of young or old plasma in the downstream of the shock.

\begin{table*}
\begin{center}
\caption{Integrated radio spectrum of the `Sausage' and `Toothbrush' relics using comparable interferometric datasets. We list integrated flux densities, errors of the integrated flux densities (see equation \ref{eq:errF}) and the noise in the images. We use these measurements to probe emission up to $\sim2.6$ arcmin on the sky. }
\vspace{-10pt}
\begin{tabular}{l r r r r r  r r r r r r r r r  }
\hline
\hline
Freq (GHz) & 0.15 & 0.24 & 0.325 & 0.61 & 1.2 & 1.4 & 1.7 & 2.25 & 2.3 & 2.75 & 3.25 & 3.75 & 16 & 30 \\  \hline
\multicolumn{5}{|l|}{`Sausage'} \\ 
Flux density (mJy) &  780.4 & &  315.7 & 222.3 & 125.7 & 117.3 & 91.2 &  61.0 & 54.3 & 50.0 & 36.1 & 27.2 & 1.2 & 0.6 \\
Error (mJy) & 80.0 & & 32.4 & 22.4 & 12.6 & 11.8 & 9.2 & 3.6 & 5.6 &  2.7 &  1.9 & 2.6 & 0.5 & 0.3 \\ 
$\sigma_\mathrm{RMS}$ (mJy/beam) & $2.4$ & & $1.0$ & $0.30$ & $0.16$ & $0.13$ & $0.15$  & $0.14$ & $0.16$ & $0.14$ &  $0.06$ & $0.17$ &   $0.06$ & $0.04$ \\
\hline
\multicolumn{5}{|l|}{`Toothbrush'} \\ 
Flux density (mJy) & 3147.9	&	1466.7	&	1042.0	&	743.1	&	344.8	&	295.5	& 236.4	& & 	162.8	& & & & 10.7 &	 3.1 \\
Error (mJy) & 314.9	&	146.8	&	104.3	&	74.3	&	34.5	&	29.6	&	23.6	&	& 16.3 & & &	&	0.7	& 0.5 \\
$\sigma_\mathrm{RMS}$ (mJy/beam) & $1.0$ &  $0.89$ & $0.24$ & $0.17$ &
$0.062$ & $0.05$ & $0.05$ & & $0.10$ & & & & $0.05$ & $0.05$ \\
\hline
\end{tabular}
\label{tab:intspec}
\vspace{-10pt}
\end{center}
\end{table*}

The images (as listed in Table~\ref{tab:imaging}), produced with uniform weighting and common uv-cut (from AMI-LA and CARMA) are shown in Figures \ref{fig:mapS} and \ref{fig:mapT} for the `Sausage' and `Toothbrush' relics, respectively. Note that even in uniform weighted images, which emphasise point sources and suppress diffuse emission, the relics are clearly detected till $30$ GHz.

 We note that uniform weighting should be used when combining different interferometric data sets with non-identical uv coverages. Uniform weighting best compensates for differences in the sampling density in the uv-plane by giving visibilities a weight inversely proportional to the sampling density function, thus accounting for any potential `holes' in the uv coverage. 

\subsubsection{Obtaining flux density measurements}\label{sec:res:flux}

We measure the flux density of the relics using uniform weighted maps that sample the same scales on the sky and that are convolved to the lowest resolution available. 

All the data were set to the same flux scale (e.g. note the very good agreement between the WSRT and VLA measurements at $\sim2.3$ GHz for the `Sausage' relic).  For the GMRT and WSRT data, we adopt a flux-scale uncertainty of $10$ per cent, resulting from telescope pointing errors and imperfect calibration as shown by \citet{2004ApJ...612..974C}  and \citet{1997A&AS..124..259R}, \citet{1998A&A...336..455S}, respectively. The AMI flux scale is precise within $5$ per cent according to \citet{2011MNRAS.415.2708A}. The VLA flux scale is constrained to better than $5$ per cent\footnote{https://science.nrao.edu/facilities/vla/docs/manuals/cal/flux/referencemanual-all-pages}. Most recently, the CARMA flux scale has been shown to be accurate to within 5 per cent \citep[e.g.][]{2013ApJ...770..112P}. In the case of the interferometric images, the error $\Delta F$ on the integrated flux density $F$ is calculated as function of the flux-scale error and image noise $\sigma_\mathrm{RMS}$:
\begin{equation}
\label{eq:errF}
\Delta F  = \sqrt{(fF)^2 + N_\mathrm{beams}\sigma^2_\mathrm{RMS}},
\end{equation}
where $N_\mathrm{beams}$ is the number of beams spanned by the source  and $f$ is the flux-scale uncertainty. 

Lower frequency, high resolution measurements show that the `Sausage' and the `Toothbrush' relics are both polarised \citep{2010Sci...330..347V, 2012A&A...546A.124V}. As explained in \citet{2014MNRAS.441L..41S}, both AMI sub-arrays measure a single polarisation (I+Q). The AMI reduction pipeline assumes random polarisation for the sources, therefore assuming that AMI measures only half the power. For randomly polarised sources the flux densities coming out of the pipeline are correct, however, the flux densities of polarised sources need to be reduced because the flux is overcorrected by the pipeline. Following the method from \citet{2014MNRAS.441L..41S}, the AMI integrated flux density of the `Sausage' relic has to be decreased by $24$ per cent, compared to the single polarisation measured by AMI. In the case of the `Toothbrush', the polarisation values from \citet{2012A&A...546A.124V} were used to derive a flux density reduction of $12$ per cent. The polarisation correction takes into account the polarisation fraction, the Faraday depth and the orientation of the electric vector with respect to the shock front at the lower frequencies to predict the rotation of the angle towards 16 GHz. Note however, that even if we assumed the sources to be fully unpolarised, their flux density would not change the overall integrated spectrum fits significantly (see below and Section \ref{sec:results}).

Figures \ref{fig:areaS} and \ref{fig:areaT} display the areas used for integration for GMRT and WSRT, overlaid on a $1.2$ GHz frequency map with the uv weighting and cut and resolution used for measuring the flux densities. We should note that flux densities of the `Sausage' relic presented in this paper will be slightly different than those calculated in \citet{2014MNRAS.445.1213S} because of small differences in the area used for integration, which in the present paper follows more closely the distribution of the emission. For the VLA, AMI-LA and CARMA datasets we decided to measure the flux in boxes closely following the distribution of emission. Imperfect deconvolution/CLEANing can result in negative bowls appearing around bright sources. This effect is mostly visible in our VLA, AMI-LA and CARMA data. In order to test this effect, we took the interferometric GMRT, WSRT, VLA, AMI-LA and CARMA images with an $800\;\lambda$ cut at their native resolution and measured the flux in tight boxes, following the diffuse emission and avoiding the negative bowls. We find that the `Toothbrush' measurements for GMRT, WSRT and AMI-LA are consistent within the error bars to the values we get when convolving to the lowest resolution and using a common integration area. The CARMA measurements are a factor of $\sim1.5$ higher when we use a tight area and native resolution, than when using a common integration area. We therefore use this new measurement in producing our integrated spectrum. In the case of the `Sausage', slightly higher flux values (factors $<1.5$) are found for the GMRT data, for the highest WSRT frequency, for the highest JVLA frequency and for AMI-LA when use tight integration areas in the native resolution images. 

We measure a flux density for the entire source as well as split the sources in two parts, to test whether the integrated spectral trends we observed are dominated by one part of the sources. Even when splitting the sources in two halves, the integrated spectrum should still be characteristic for a summation of shock accelerated populations with ageing behind the shock. The `Sausage' and `Toothbrush' relic flux density measurements at high resolution from interferometric images can be found in Table~\ref{tab:intspec}. We also give the flux densities of the subsections of the relics in Table~\ref{tab:intspecsub}.

Figures~\ref{fig:intspecS} and \ref{fig:intspecT} show the flux density measurements and spectra of the `Sausage' and `Toothbrush' relic, respectively. As found by \citet{2014MNRAS.441L..41S}, a single power law provides a poor description of the data. We therefore fit the spectrum of each relic with two power laws, one using the measurements below $2.5$ GHz, and the other using the measurements above $2.0$ GHz. We use weighted least-squares regression, where each point is down-weighed according to its error (as per equation \ref{eq:errF}). We fit the flux density for the entire source, as well as split the source in two regions (left and right half for the `Sausage' relic and the subareas B1 and B2+B3, see Figure \ref{fig:mapT}). The measurements and the results of the fit can be found in Figures \ref{fig:intspecS} and \ref{fig:intspecT}, for the `Sausage' and `Toothbrush' relics respectively. We find that the low-frequency spectrum ($<2.5$ GHz) greatly differs from the high frequency spectrum ($>2.0$ GHz) for both sources. The low frequency spectrum for the `Sausage' relic has a spectral index of $\alpha=-0.90\pm0.04$, while beyond $2.5$ GHz this steepens to  $-1.71\pm0.11$ (difference significant at the $6.4\sigma$ level). In the `Toothbrush', the low frequency spectrum differs from the high frequency one at the $>5.8\sigma$ level ($-1.00\pm0.04$ versus $-1.45\pm0.06$). The results hold when splitting the source into two subareas, as explained above. Note than when we use fluxes measured in the same box at all frequencies, we obtain spectra with slopes consistent within the error bars with the slopes we obtain by using the fluxes obtained by best following the emission.

\subsection{Total power measurements}
Radio relics are diffuse objects, so any spatial filtering applied by interferometers will affect the measured total flux density. We also attempt to create a spectrum with our interferometric data that have good short baseline coverage. This enables us to approximate a total power measurement from interferometric dataset and compare them with Effelsberg data. For this purpose we can use the GMRT data as well at the lowest frequency WSRT measurements which have baselines down to 100 $\lambda$ corresponding to a largest spatial scale probed of $\sim21$ arcmin. The rest of our interferometric data cannot properly pick up the largest scales of the relic emission, which Effelsberg detects.

\subsubsection{Effelsberg data}\label{sec:effdata}

The low resolution, total power images from Effelsberg are shown in Figures \ref{fig:mapS} and \ref{fig:mapT} for the `Sausage' and `Toothbrush' relics, respectively. In the low resolution images, the relics are detected at very high $S/N > 16$, but isolating the relic emission from the point sources is non-trivial. 

Several radio galaxies are located nearby the relics which, could blend with the diffuse emission, depending on the resolution. In the interferometric images (with higher resolution, $\sim 40$ arcsec), blending is not a contaminant, since no radio sources are located too close to the diffuse emission. This can be clearly seen in very high resolution ($\sim 5$ arcsec), deep ($\sim 25$ $\mu$Jy) GMRT images of the two relics \citep{2012A&A...546A.124V,2013A&A...555A.110S}.

However, contamination by radio galaxies is an important effect for the Effelsberg data. Radio galaxies are expected to have a flatter \citep[$\alpha \sim -0.7$;][]{1992ARA&A..30..575C} integrated index compared to the radio relics, and will, if anything, bias high the flux density measurement of the relics at higher frequencies. If the radio galaxy contribution is not properly subtracted, the relic flux density can be overestimated. The contamination is most important for the `Sausage' relic, which is neighboured by an unrelated radio AGN (source H, see Figure~\ref{fig:mapS}) towards its eastern edge, which has a spectral index of $-0.77 \pm 0.04$ (as measured from the interferometric images the with $800\,\lambda$ inner uv cut). 

The integrated flux densities for Effelsberg were measured using the option `tvstat' in \textsc{AIPS}. For both relics and at both frequencies we defined integration areas around the relic down to a level where the intensity reaches the noise. The point source flux densities are calculated by interpolating between their fluxes in the higher resolution images and subtracted from the relic flux density. For the `Sausage' relic at $4.85$ GHz, we subtract sources B and H (see Figure~\ref{fig:mapS}) from the flux density value by measuring their fluxes in higher resolution VLA C-array data at $4.9$ GHz \citep[data presented in][]{2010Sci...330..347V}. At $8.35$ GHz, source H was subtracted, as source B could be avoided for the integration. For the `Toothbrush' relic, we subtract point sources F and G, using their fluxes measured in the $800\,\lambda$ cut intereferometric measurements at $2.3$, $16$ and $30$ GHz. Note however, the extrapolation of the flux density of contaminating radio sources is imperfect and could result in an overestimation of the relic flux density by $5-10$ per cent. 

The uncertainties of the Effelsberg flux density measurements are dominated by the uncertainties in the baselevel of the maps. To correct the baselevel of each final map in Stokes I, Q and U, the mean intensity of each map was measured by selecting at least five boxes in regions where  little to no emission from sources was detected. The average of these values was subtracted from the final map to result in maps with a baselevel near to zero. In the case of the `Sausage' $\sim0.03$ mJy/beam (at $8.35$ GHz where the noise is $\sim0.07$ mJy/beam) and $\sim0.11$ mJy/beam (at $4.85$ GHz, where the rms noise is ~0.12 mJy/beam) were subtracted from the baselevel. In the case of the `Toothbrush', $\sim0.37$mJy/beam (at $8.35$ GHz, where the rms noise is ~0.13 mJy/beam) and $\sim0.53$ mJy/beam (at $4.85$ GHz, where the rms noise is ~0.10 mJy/beam) were subtracted. However there still is a possibility that the Effelsberg measurements do not lie on the same flux level as the other interferometric data we aim to compare them with. The Effelsberg maps are too small to find large enough regions without signal: because of the large beams, the diffuse and compact flux persists in the edges of the maps. Therefore the uncertainties in the baselevel corrections are quite large. However, the diffuse emission is very well confined at the northern edge of the relics: this is clearly visible from higher resolution data. There is no diffuse cluster flux at the north, east of west of either relic \citep{2012A&A...546A.124V, 2013A&A...555A.110S}. The only diffuse emission emission, at very low level in both clusters, is located directly south of the relics where there are hints for a giant radio halo. However no baselevel measurement boxes were placed in those areas. Therefore our method of setting the baselevel does not result in a subtraction of the diffuse component from the Effelsberg measurements. 

The error on the Effelsberg measurements is:
\begin{equation}
\label{eq:errFEff}
\Delta F  = N_\mathrm{beams}\sigma_\mathrm{b},
\end{equation}
where $N_\mathrm{beams}$ is the number of beams spanned by the source and $\sigma_\mathrm{b}$ is the uncertainty of the baselevel (the standard deviation of the mean intensity in the boxes where the baselevel was measured). The errors on the Effelsberg flux densities could be in reality larger if we would account for any point source contamination remaining after the point source subtraction (which could be of up to 5-10 per cent). We also note that the baselevel correction is large for the `Toothbrush' data, higher than the image rms noise. Therefore there could be additional systematics that are not included in the error bar.

The Effelsberg calibration was compared to the interferometric $4.9$ VLA images of the `Sausage' field \citep{2010Sci...330..347V}, convolved to $90$ arcsec resolution. The flux densities of sources B and H (see Figure~\ref{fig:mapS}) are consistent with $-0.82$ and $-0.87$ spectral indices between $4.9$ GHz (VLA) and $8.35$ GHz (Effelsberg). 

\begin{figure}
\begin{center}
\includegraphics[trim=0cm 0cm 0cm 0cm, width=0.495\textwidth]{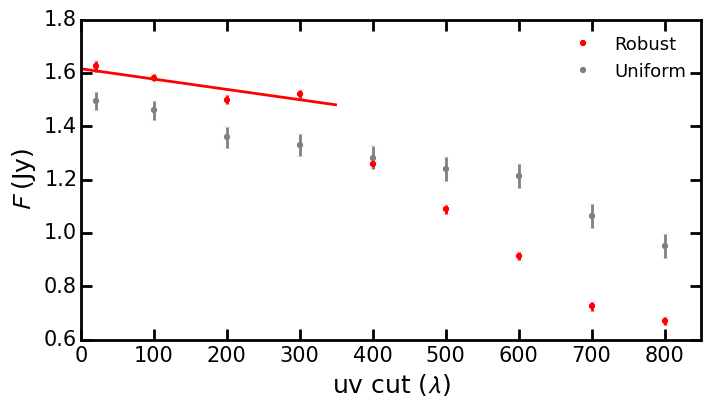}
\end{center}
\vspace{-10pt}
\caption{Dependence of the measured integrated relic flux density on the weighting and uv cut employed. Note that the uniform weighting is not as sensitive to  uv cut as the robust weighting. We use a fit to the points with uv cuts smaller than 300 $\lambda$ to predict total power measurements for comparison with the Effelsberg data.}
\label{fig:uvcuts}
\vspace{-10pt}
\end{figure}

\subsubsection{Obtaining a total power measurement from interferometric measurements}\label{sec:interftotpower}

We image the two relics at native resolution using the full uv coverage for GMRT and WSRT 1.2 GHz. We employ robust weighting to maximise the detection of large scale flux. The GMRT datasets at 150, 240 325 and 610 MHz have baselines down to 20, 40, 70 and 120 $\lambda$, respectively, while the WSRT 1.2 GHz dataset goes down to 100 $\lambda$ uv distances. 

The interferometric images, even imaged to best emulate a total power measurement, are not fully comparable to the Effelsberg observations, given interferometers resolve out flux detected by the single dish at Effelsberg. To assess how much flux can potentially be lost with interferometric imaging, we use the 150 MHz data for the `Sausage' field, which has the best inner uv coverage going down to $20\,\lambda$. The 150 MHz data set is suitable for this purpose since it covers spatial scales that are even larger than the cluster itself and has dense inner uv coverage. We image the data with different inner uv cuts, using both robust (with robust factor set to 0.5) and uniform weighting. As explained in Section~\ref{sec:res:images}, uniform is the least sensitive weighting to the actual uv distribution. The robust parameter is routinely set to 0.5 for displaying our radio images. This enhances the diffuse emission, without losing too much resolution. However, this robust weighting leads to images and flux measurements that can be sensitive to the actual distribution of the uv data points. Hence images made with robust weighting are not used to derive the radio spectrum. Figure~\ref{fig:uvcuts} shows a plot of the integrated `Sausage' flux density, measured in the same region, versus the minimum uv distance used for imaging. This plot is helpful in roughly estimating how much flux is lost depending on the inner uv cut. Note that these trends are confirmed by our simulated radio observations in Section~\ref{sec:lowSN}. Note than the uniform weighting is not as affected as the robust weighting by the inner uv cut, as expected. For example, the `Sausage' relic $150$ MHz flux density drops by a factor of $\sim1.6$ when using baselines beyond $800$ $\lambda$ instead of beyond $20$ $\lambda$ and uniform weighting. Therefore, we expect Effelsberg to measure a higher flux density than inferred from the $>800$ $\lambda$ interferometric observations only (such as those shown in Figures~\ref{fig:intspecS} and \ref{fig:intspecT}, see also Section~\ref{sec:effdata}). The Effelsberg values are indeed higher or consistent with the fluxes predicted from the best fit lines from Figures~\ref{fig:intspecS} and \ref{fig:intspecT}.

We fit the dependence of the relic flux density on the uv cut (using the uv cuts lower than 300 $\lambda$) and use this linear fit to predict total power measurements from the interferometric images. Note that the correction is imperfect given that structures at different frequencies and with different uv coverages will be recovered differently. However, we expect the flux dependence on the uv cut from Figure~\ref{fig:uvcuts} to be characteristic for all frequencies in the case of these two relics. As explained in Section~\ref{sec:res:images}, because the `Sausage' and `Toothbrush' relics are highly elongated in the east-west direction, the flux lost through the uv cut will be aligned with the shock structure and the downstream area of the shock is fully probed. Therefore, the uv cut will not bias the spectral shape (this is confirmed in Section~\ref{sec:res:totalflux}). Additionally, the quality of the 150 MHz dataset are quite similar to the other frequencies. The detection significance at low and high frequency are quite similar, in the range of $S/N\sim10$. The `Sausage' relic is detected at low S/N ($\sim2-3$) only in the AMI and CARMA data, but very well detected in the VLA datasets ($S/N\sim20$), where the steepening is already visible. In the case of the `Toothbrush', the relic is detected at $S/N\sim15$ in AMI and $S/N\sim6$ in CARMA.

We measure the GMRT and WSRT radio relic flux densities in the native resolution of the images  to compare with Effelsberg measurements. If we were to smooth to the Effelsberg resolution, the flux densities would increase by a factor of at least $1.5$. However, a correction for the contamination by radio sources would have to be included, which is imprecise as extrapolations need to be made. The flux density of radio galaxies can increase by a factor of up to $1.3$, from a resolved to a heavily unresolved morphology. We tested the method of measuring the high resolution point source flux, smoothing the image and then subtracting the point source contamination but the results were within the error bars. Note that given the different resolutions of the interferometric and single dish data make it impossible for the flux densities to be measured in the same regions.

We calculate the error on the interferometric flux densities as in equation \ref{eq:errF}, however, we add 5 per cent to the flux-scale uncertainty to account for the different integration areas used in the low frequency maps and the Effelsberg measurements as well as to account for the extrapolation we make to total power flux densities. 

\subsubsection{Total power spectrum}\label{sec:res:totalflux}

\begin{figure}
\begin{center}
\includegraphics[trim=0cm 0cm 0cm 0cm, width=0.495\textwidth]{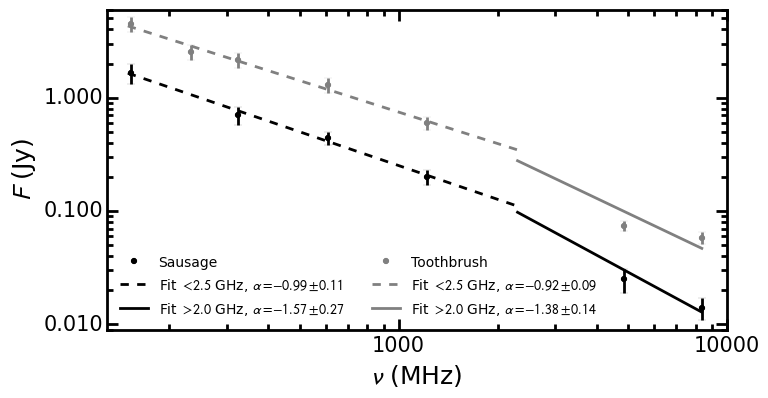}
\end{center}
\vspace{-10pt}
\caption{Integrated `Sausage' and `Toothbrush' relic spectrum using the Effelsberg measurements and interferometric observations corrected to obtain a total-power value.  We correct the interferometric measurements using the predictions from Figure~\ref{fig:uvcuts}, with factors of $1.1-1.17$ depending on the inner uv coverage at each frequency. Note the values of the interferometric measurements are not fully comparable with the total power Effelsberg measurements, but this is included in the error bars. Note the spectrum is steepening towards high frequencies as observed from spectra made using comparable interferometric datasets probing the same spatial scales. A 2.0 GHz measurement is predicted using the low-frequency fit. This prediction is then combined with the Effelsberg flux densities to obtain a high frequency spectrum.}
\label{fig:totalpower}
\vspace{-10pt}
\end{figure}

The Effelsberg total power flux density measurements and the GMRT and WSRT fluxes emulating a total power measurement can be found in Figure~\ref{fig:totalpower} and Table~\ref{tab:intspecTP}. We present the total power spectrum for reference only, given that the measurements are not fully comparable. Our main goal is to see whether to total power measurements are consistent with the results found from the interferometric spectrum. We fit the flux density measurements below 1.2 GHz with a power law and find that the slopes are  indeed consistent within the error bars with the slopes found from the higher resolution maps with 800 $\lambda$ cut. This indicates that applying an inner uv cut does not bias the spectral shape. For the `Sausage' relic we find a slope of $-0.99\pm0.11$ (compared to $-0.90\pm0.04$ in the high resolution images), while for the `Toothbrush' we obtain a slope of $-0.92\pm0.09$ (versus $-1.00\pm0.04$ in the high resolution maps). We use these fits to predict flux densities at $2.3$ GHz and combine this prediction to obtain a high frequency slope, comparable with the fits in Section \ref{sec:res:flux}. We assume an error bar on the $2.3$ GHz comparable with the 1.2 GHz error. In the case of the `Sausage' relic, as found in Section \ref{sec:res:flux}, the high frequency spectrum steepens with respect to the low frequency one with a slope of $-1.57\pm0.27$, consistent with the steepening found in the high resolution maps ($-1.71\pm0.11$). The `Toothbrush' relic high frequency, total power measurements indicate a spectral index of $-1.38\pm0.14$ (versus $-1.45\pm0.06$). Therefore the low-resolution, total power measurements confirm the steepening found in Section \ref{sec:res:flux}. Note however, that because of a lack of data points within the $1-4$ GHz regime and the large errors on all the measurements, the total power spectral fit is not as well constrained as in the case of the interferometric spectrum. Because of the large extrapolation made for the measurements coming from interferometers, we employed large, conservative errors (as discussed at the end of Section \ref{sec:interftotpower}). In the case of the `Toothbrush', a single power law can be rejected at the $99.9$ per cent confidence level. In the case of the `Sausage', we cannot exclude a single power law fit based on the total power data alone. Note however, that the reduced $\chi^2_\mathrm{red}$ of about $0.3-0.4$ for the low frequency fit of both relics Figure \ref{fig:totalpower} indicates that we are likely overestimating our errors at the low frequency part. We test the effect of making the errors smaller to the point where a low frequency single power law fit gives us a reduced $\chi^2_\mathrm{red}$ of about 1 (where the errors should reflect the spread in the data). If we were to assume these smaller errors on the low-frequency data points, the single power law fit is rejected for both relics at the $99.9$ per cent confidence level.

\section{Discussion}

\subsection{Possible systematic effects decreasing high frequency flux densities}\label{sec:disc:syst}

We explore possible systematic, instrumental effects which can cause the high frequency measurements to be biased low. We refer to the spectrum produced using comparable interferometric measurements as the interferometric spectrum and to the spectrum produced from Effelsberg and the interferometric data with best inner uv coverage as the total-power spectrum.

\subsubsection{Mosaicking in the interferometric spectrum}
The VLA, AMI-LA and CARMA data are all from mosaics, which means the primary beam correction might be imperfect, especially if the pointings are separated at large distances. AMI-LA has a primary beam of about $5.5$ arcmin, which in principle means two pointings can cover the `Sausage' and `Toothbrush' relic over they full extent. However four pointings over the relics were used to alleviate the issue of the imprecise primary beam correction at large large distances from the pointing centre. Similarly, the CARMA observations were designed with $11$ pointings over the sources. The VLA observations are also not susceptible to such errors, where two, closely spaced pointings were used to cover the `Sausage' relic. Even at the highest VLA frequency, the radio relic is located in areas where the beam level is over 70 per cent of the peak sensitivity. At such levels, the primary beam is very well known and primary beam uncertainties are very small ($<<1$ per cent). We further test the impact of the primary beams in the context of mosaicking, by measuring the VLA spectra of radio galaxies in the field \citep[sources A, B, C, D, E, H using the labelling from][and unlabelled source located north of source H]{2013A&A...555A.110S}. The spectral indices of the these source vary between $-1.26$ and $-0.93$, with no systematic trends across the field of view (see also Figure~\ref{fig:pointsources}. This indicates that primary beam uncertainties do not affect the measurements.

\begin{table}
\begin{center}
\caption{Integrated radio spectrum of the `Sausage' and `Toothbrush' relics using the Effelsberg single dish and interferometric measurements best simulating a total-power measurement. We give the integrated flux densities and their errors. Single dish measurements are given with point sources subtracted.}
\vspace{-10pt}
\begin{tabular}{l r r r r r r r }
\hline
\hline
Freq (GHz) & 0.15 & 0.24 & 0.325 & 0.61 & 1.2 & 4.85 & 8.35 \\  \hline
\multicolumn{5}{|l|}{`Sausage'} \\ 
Flux density (mJy) &  1655 &  & 700 & 440 & 200 & 25 & 14 \\
Error (mJy) & 337 & & 126 & 60 & 30 & 6 & 3 \\ 
\hline
\multicolumn{5}{|l|}{`Toothbrush'} \\ 
Flux density (mJy) & 4470 &	 2510 & 2150 &	1290 & 600 & 74 & 58  \\
Error (mJy) & 680 & 380 & 320 & 190 & 80 & 8 & 7 \\
\hline
\end{tabular}
\label{tab:intspecTP}
\vspace{-10pt}
\end{center}
\end{table}

\subsubsection{Working in low S/N regimes}\label{sec:lowSN}

\begin{figure}
\begin{center}
\includegraphics[trim=0cm 0cm 0cm 0cm, width=0.495\textwidth]{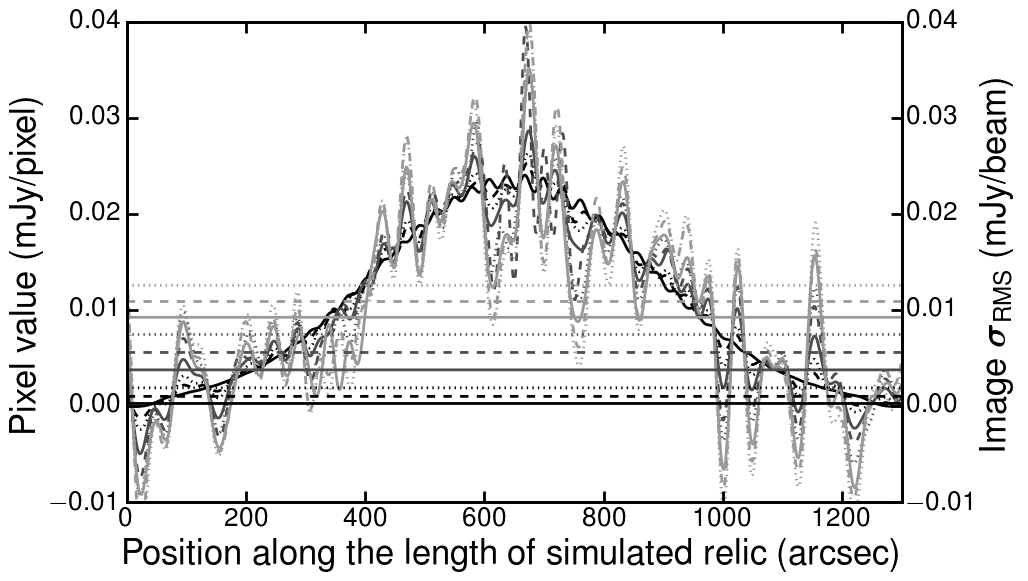}
\end{center}
\vspace{-10pt}
\caption{East-west cut through a simulated `Sausage'-like CARMA observation with increasing noise levels (curved lines). The horizontal lines indicate the image $\sigma_\mathrm{RMS}$ in the same line-style and colour.}
\label{fig:simulation}
\vspace{-10pt}
\end{figure}

\begin{figure}
\begin{center}
\includegraphics[trim=0cm 0cm 0cm 0cm, width=0.495\textwidth]{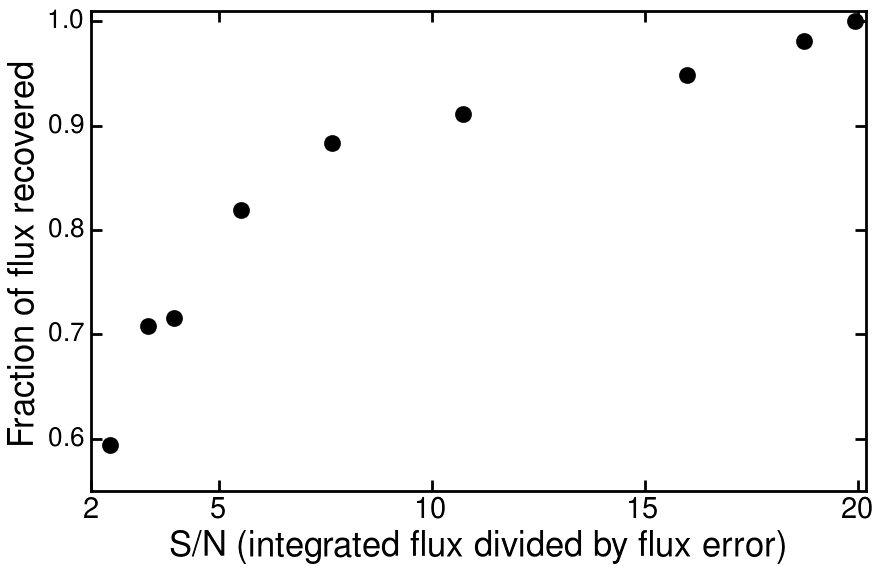}
\end{center}
\vspace{-10pt}
\caption{Flux recovered through CLEANing in a simulated `Sausage'-like CARMA observation with increasing noise levels (curved lines). The amount of flux recovered decreases with S/N (defined as integrated flux divided by the flux error). The flux is measured by integrating the area over which the brightness is above 0 in the no noise, reference simulation.}
\label{fig:recovery}
\vspace{-10pt}
\end{figure}

 In some of our maps, the diffuse emission is detected at low S/N, more importantly in the AMI and CARMA image of the `Sausage' where the emission throughout the relic is detected at the $3-10\sigma_\mathrm{RMS}$ level. Deconvolution is a non-linear process and attempts to interpolate/extrapolate over holes/gaps in the uv-plane. Therefore the performance of  CLEANing can depend on the S/N. In order to test this, we produce a simple mock radio observation designed to mimic our CARMA observations of the `Sausage' relic. We model the `Sausage' relic at 30 GHz as a Gaussian, highly elongated in the east-west direction, with a major of axis of 9 arcmin and minor axis of 30 arcsec. We set the flux density of the source to $0.75$ mJy and the sky coordinates similar to the real `Sausage' relic. We use the `simobserve' task in \textsc{casa}, version 4.3, which produces mock visibilities that would be measured by observing the `Sausage' sky brightness model. We use the actual CARMA telescope configuration, the same observing time and the mosaic mode, placing the pointings at 1.4 arcmin distance, as in the real CARMA observation. We produce a range of mock radio observations. The reference dataset has no noise in the visibilities, but we add a range of Gaussian noise values to better simulate real observing conditions. In this way we produce mock observations which have identical uv-coverage, but have increasing image noise. This simulates a scenario with decreasing S/N for the source of interest. We image the mock uv datasets using the `simanalyze' function in \textsc{casa}, in which we use CLEANing in the same way we did for the real observations, specifically using uniform weighting and  CLEAN boxes. Figure~\ref{fig:simulation} shows a east-west cut through the CLEANed images with increasing noise. Figure~\ref{fig:recovery} indicates the amount of flux density recovered as function of S/N (defined as integrated flux recovered divided by the error on the flux). The error is calculated as in equation \ref{eq:errF}. The horizontal line indicate the $\sigma_\mathrm{RMS}$ in the maps. There are two main outcomes of the simulations:

1. Flux is lost in the east-west direction, even in the simulation with no noise, because of the inner uv cut imposed by the CARMA uv distribution. The flux density is reduced from 0.75 to 0.6 mJy, in agreement with the results presented in Figure~\ref{fig:uvcuts} using real data.

2. Throughout all the images, irrespective of the noise level, the overall brightness distribution is very similar to the reference case with no noise. Note however, that the real brightness distribution of the relics in the CARMA and AMI observations are slightly different than in our simulation: the brightness distribution is more uniform across the length of the relic, resulting a approximately constant $2-10\sigma$ peak values across the source. We measure the flux density by integrating over the area over which the simulated relic can be traced above 0 in the reference case. As expected, the flux recovered decreases with S/N (Figure~\ref{fig:recovery}). 

Our results are in perfect agreement with the simulations of flux recovery for diffuse sources by \citet{2003ApJS..145..259H}. The real CARMA and AMI observations are comparable in terms of per-pixel and integrated S/N to a flux recovery of $65-85$ per cent. This means that the real flux (after the uv cut effect) could be $1.15-1.55$ times higher than what CLEAN recovers. Note that these allowed values are within $1-1.5\sigma$ of the values we measure, therefore not in tension with our observed measurements. Note however, that even a factor of 1.5 increase in CARMA and AMI `Sausage' flux does not significantly change the high frequency fits we present in Section~\ref{sec:res:flux}. 

Therefore we conclude that indeed in our AMI and CARMA `Sausage' maps we are affected by imperfect CLEANing behaviour because of low S/N, but the effect is too small to explain the high frequency steepening we observe in the integrated relic spectrum.

\subsubsection{Flux calibration}

\begin{figure}
\begin{center}
\includegraphics[trim=0cm 0cm 0cm 0cm, width=0.495\textwidth]{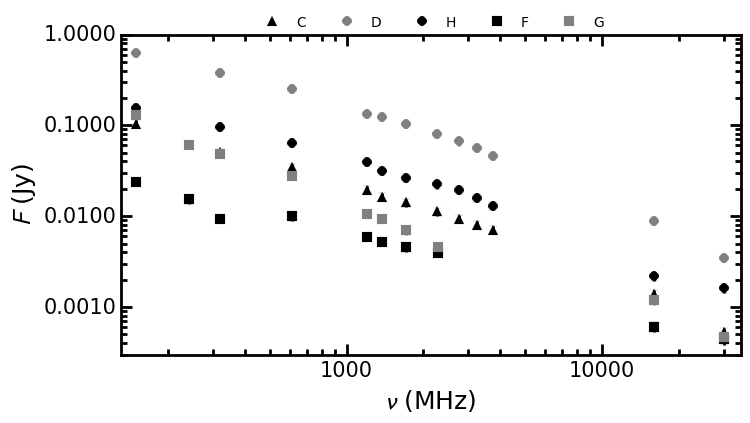}
\end{center}
\vspace{-10pt}
\caption{Integrated spectra for radio galaxies in the `Sausage' (circles) and `Toothbrush' (squares) fields. Source with clear detections at all frequencies are plotted. For labelling see Figures~\ref{fig:mapS} and \ref{fig:intspecT}. The spectra show that there are no systematic flux-scale errors or primary beam imperfections affecting our data.}
\label{fig:pointsources}
\vspace{-10pt}
\end{figure}

An issue with the flux scale at a particular frequency could result in an  over or underestimation of the flux. The spectra for bright sources well detected at all frequencies are shown in Figure~\ref{fig:pointsources}. The integrated spectra of these radio galaxies n the GMRT, WSRT, VLA, AMI-LA and CARMA data are described by smooth curves, indicating all the observations are on the same flux scale. Note that at higher resolution, galaxies C, D, H in the `Sausage' field and F in the `Toothbrush' field' are actually old, head-tail radio sources where ageing of electrons in the lobes can be clearly detected \citep{2012A&A...546A.124V, 2013A&A...555A.110S}. The integrated spectrum of such sources is expected to slightly steepen at high frequencies because of synchrotron losses \citep{1992ARA&A..30..575C}. However, there are no systematic breaks happening at a particular frequency, which could indicate a particular observation being on the wrong flux scale.  

Also note the very good agreement between the WSRT and VLA measurements at $\sim2.3$ GHz for the `Sausage' relic, indicating that the WSRT and VLA data are on the same flux scale. Moreover, we are finding a steepening in the spectrum of the two relics using data taken with three independent telescopes. It is highly unlikely that the flux scale of AMI, CARMA and VLA is biased low for independent measurements of two sources. 

We therefore conclude that flux-scale issues cannot explain the steepening of the relic spectrum at high frequencies.

\subsubsection{Sunyaev-Zeldovich decrement}
 
At frequencies below $\sim217$ GHz, the SZE is detected in decrement relative to the baseline given by the cosmic microwave background, causing a reduction in the measured flux densities of sources close to the cluster center, where the SZE is strongest. The SZE is a cluster-wide, large-scale effect.  However, at the frequencies we are probing in this paper, the SZE decrement is expected to be small even at the cluster centre. For example, the Coma cluster hosts a giant diffuse radio halo which follows the ICM X-ray distribution. The high frequency steepening of the halo spectrum was originally attributed to the SZE decrement. \citet{2010MNRAS.401...47D} and \citet
{2013A&A...558A..52B} showed that the decrement integrated over the entire cluster is a factor of $4-5$ too small to explain the steepening observed in single dish measurements. Additionally, at the resolution of AMI-LA and CARMA the largest scale detectable is limited by virtue of interferometric observations. For example, AMI-LA observations of the clusters in the XMM Cluster Survey \citep[XCS;][]{2013MNRAS.433.2920A} and the Local Cluster Substructure Survey \citep[LoCuSS;][]{2012MNRAS.425..162A} resulted in no SZE detection. Additionally the AMI-SA observations measured a total SZE flux $<2$ mJy, indicating the SZE contribution will be extremely small ($\lesssim 50$ $\mu$ Jy, or less than $\sim5$ per cent of the relic flux density) at AMI-LA resolution and at the large cluster-centric distances the `Sausage' and `Toothbrush' relics are located ($1.5$ Mpc). Even though the relics are located far away from the cluster centre, \citet{2015MNRAS.447.2497E} argue that shock compression could boost the SZE signal, leading to up to $30$ per cent decrease at $10$ GHz for Effelsberg observations of the relic of Abell 2256 \citep{2015A&A...575A..45T}. There is an important difference between our observations and Abell 2256: the relic in Abell 2256 is about 500 kpc from the cluster center, while the `Sausage' and `Toothbrush' clusters are $1.5$ Mpc away from the cluster centre. \citet{2015A&A...575A..45T} argue that in the case of Abell 2256 the SZE decrement is of the order of $30$ per cent at the distance of the relic from the cluster center (about $500$ kpc). They estimate that at a distance of $750$ kpc the SZE induced decrement drops to $15$ per cent. However, the `Sausage' and `Toothbrush' relics are located about $1.5$ Mpc away from the cluster centre, hence the SZE contamination would be very small. A linear extrapolation from the estimations of \citet{2015A&A...575A..45T} indicates a contamination at the level of $<5$ per cent, while a beta-profile for the ICM density would lead to even smaller numbers. Note that even if a significant decrement of a factor of $2$ would be contaminating the AMI, CARMA or Effelsberg relic measurements, that would still not be enough to explain the discrepancy between the expected flux densities by extrapolation from the low frequency measurements and the measured fluxes. By extrapolating the low frequency spectrum of the `Toothbrush', the AMI and CARMA flux densities would have to be $3.5$ and $6$ times higher than measured, respectively. In the case of the `Sausage' the flux densities predicted from the low frequency spectrum are more than $10$ times higher that those measured.

\subsection{Why is the integrated relic spectrum steepening?}\label{sec:results}

Using measurements at $\sim15$ independent frequencies we find that the high frequency spectrum of the `Sausage' and `Toothbrush' relics differs significantly ($\gtrsim6\sigma$) from the low frequency spectrum. Our estimates indicate systematic biases cannot account for the observed measurements. 

In the simple radio relic formation scenario proposed by \citet{1998A&A...332..395E} the shock strength is constant in time, the magnetic field is constant across the source and thermal particles are injected at the same rate. Under these assumptions, the index of the integrated spectrum should be $0.5$ steeper than the injection index \citep{1998A&A...332..395E} and the injection index $\alpha_\mathrm{inj}$ can be directly translated to a Mach number through $M=\sqrt{(2\alpha_\mathrm{inj}+3)/(2\alpha_\mathrm{inj}+1)}$ \citep[for a simple, non-relativistic hydrodynamical shock with a weak magnetic field;][]{1987PhR...154....1B}. This only holds if we are observing the spectrum away from the break frequency ($\nu_\mathrm{br}$), where we detect the electrons which have started to lose energy through synchrotron and inverse Compton processes at a steady rate. Our measurements are incompatible with this simple model currently adopted for the formation of radio relics and explore other scenarios that could result in a curved integrated spectrum. If any of the assumptions of the model are broken, the integrated spectral steepening result can differ from the canonical $0.5$.

We now discuss possible physical scenarios which would lead to a curved integrated relic spectrum. In the following discussion we focus on the spectra produced from comparable, high resolution ($\sim40$ arcsec) interferometric maps produced with an $800\,\lambda$ inner uv cut.

\subsubsection{Non stationary shock conditions}
Simulations by \citet{2015JKAS...48....9K} show that stationary shock conditions \citep[as in the model of][]{1998A&A...332..395E} cannot be assumed for cluster merger shocks as they are expanding into a medium with decreasing density and temperature. Our data supports these simulations, since the integrated spectral index of the two relics is about $-0.9$ at low-frequency and steepening to more than $\sim-1.5$ at higher frequencies. If we assume we are observing far away from $\nu_\mathrm{br}$, this is inconsistent with the simple \citet{1998A&A...332..395E} model, as subtracting $0.5$ would result in an injection index that is too flat and an infinite Mach number. 

\subsubsection{Structure in the Mach number or magnetic field distribution}
If the magnetic field or Mach number are not uniform across the source causing anisotropies in the rate of particles injected, then the assumptions of the simple DSA scenario proposed by \citet{1998A&A...332..395E} are broken. The Mach number for example would mildly strengthen as it propagates outwards, impinging on lower density and lower temperature ICM gas \citep{2014IJMPD..2330007B}. However, observations of the brightness distribution, spectral index and electron age of the two relics indicate a relatively uniform structure, and hence a  relatively constant Mach number throughout the source. Therefore anisotropies in the Mach number are not a pivotal factor in shaping the relic spectrum \citep{2010Sci...330..347V, 2012A&A...546A.124V, 2013A&A...555A.110S, 2014MNRAS.445.1213S}. 

Simulations indicate that magnetic fields at the location of the shock are aligned and amplified \citep{2012MNRAS.423.2781I}, while significant turbulence develops in the shock downstream area \citep{2011ApJ...726...17P}. Note however, such effects are heavily dependent on the assumed model. A slowly decaying magnetic field from the shock region into the downstream area would lead to a decreasing cooling rate of the electron away from the shock. 

A magnetic field stronger at the shock than in the downstream area would lead to an integrated spectrum dominated by the freshly accelerated electrons located in the immediate vicinity of the shock front. The spectrum will therefore be closer to what is expected for freshly DSA-accelerated electrons, so not $0.5$ steeper spectrum as usually assumed in the case of a uniform mix of fresh and ages electron. If one would assume this integrated spectrum come from continuous injection of electrons, and subtract $0.5$ to obtain an injection spectrum, the injection index derived will be too flat and the Mach number overestimated. However, the integrated spectra would depend on how the magnetic field varies (as the energy of electrons emitting at a given frequency changes as function of magnetic field). For instance, if the field is strong at the shock the high energy particles radiate their energy quickly. A field weakening into the downstream area means that the electrons needed to emit at a given frequency will disappear more quickly, so that the spectrum is actually steepened more. In order to reproduce a curving spectrum, we would need to amplify cooling in the downstream area. 

The turbulence in the downstream area might re-accelerate the particles injected by the shock, instead of suppressing the high frequency emission. Any downstream re-acceleration would be expected to boost the brightness of emission of aged electrons, most likely leading to a very flat spectrum, rather than a curved one. Additionally, turbulent re-acceleration is a slow process as it requires the large bulk motions to cascade into small scale turbulence ($>>100$ Myr) which are able to accelerate particles with sufficient efficiency \citep{2007MNRAS.378..245B}.

\subsubsection{Seed population with energy distribution cut-off} 

Another scenario would be shock re-acceleration of a seed population with a cut-off in the energy distribution. Including both locally injected, thermal particles and particles coming from upstream area of the shock, DSA takes all the particles in a small energy range and redistributes them into a power law at higher energies with a certain slope \citep{1983RPPh...46..973D}. If all the particles enter the process at energies well below ($4-5$ orders of magnitude) the one of interest then the appropriate slope at the energy of interest is the `standard' test particle slope \citep{1993PASAu..10..222M}. If, on the other hand, the initial population being accelerated includes particles around the energy of interest ($\gamma\sim10^4$), the relative contributions starting at different energies have to be considered, resulting from a blend of particles that entered at different starting energies which need to be weighted accordingly \citep[e.g.][]{2000A&A...357.1133G}. To a first approximation, if the initial population at the energy of interest has a steeper spectrum than the classic DSA test particle slope, the outcome is the classic slope. For example, \citet{2011JKAS...44...49K} and \citet{2011ApJ...734...18K} show that when pre-accelerated particles are considered, the overall spectral shape is conserved, as long as the seed population does not have a spectral break and the spectrum is flatter than the injection index given by the Mach number of the shock (hence the population is young, recently accelerated). Therefore, only the normalisation changes, because of the more efficient particle injection. Therefore, the DSA acceleration scenario, when modified for young relativistic particles, cannot explain the steepening of the spectrum towards high frequencies. 

If, instead, the initial spectrum at or just below the one of interest is flatter than the classic slope for that shock, the outcome slope is the same as the one coming in \citep[e.g.][]{2011ApJ...734...18K}. This was also observed in simulations tailored to the `Sausage' relic, which find a good agreement to the observed low-frequency ($<1.5$ GHz) spectrum \citep{2012ApJ...756...97K}.

If the incoming spectrum steepens or cuts off at high energies, then the output spectrum at high energies will take the flatter value of that coming in and the classic DSA slope for this shock. DSA always adds energy to the particle population \citep{1983RPPh...46..973D}. Therefore, at a given energy, the number of particles after DSA acceleration will be  higher than the number of particles at that energy before DSA \citep{1983RPPh...46..973D}. Every particle gains, on average, an amount of energy proportional to its initial energy with a factor that depends on the number of times it bounces across the shock and the shock speed jump. The efficiency of energy transfer in DSA can be larger when an upstream, pre-accelerated population is involved, since the energy gain scales with the initial energy. 

A few studies attempted to model radio relic spectra using DSA re-acceleration of seed, relativistic electrons with a break in the energy spectrum. Most recently, \citet{2015arXiv150504256K} performed DSA simulations matched to the `Sausage' relic, which indicate that a seed population with a break in the energy distribution, once re-accelerated by the shock, would result in a steepening integrated spectrum. The authors suggest that electrons sourced from past AGN activity or previous shocks, have since aged, resulting in a distribution with an exponential fall-off a high energies. Fossil electrons previously accelerated by accretion shocks would reside at a few $100$ MeV energies and have a very long lifetime of a few Gyr, without being directly detectable in the radio. Observationally, this is supported by studies such as \citet{2015MNRAS.449.1486S} where an active radio galaxy is most likely feeding the plasma for a nearby relic. The authors suggest that pre-accelerated plasma could stream for Mpc distances along lines of equipotential ICM specific entropy, which at large cluster-centric distance can be approximated by spherical shells. However, none of the scenarios \citet{2015arXiv150504256K} tested fully matched the GMRT, WSRT and AMI-LA observations from \citet{2014MNRAS.441L..41S} of the `Sausage' cluster, and the authors conclude additional processes apart from radiative losses might be operating in radio relics. 

However, \citet{2015arXiv150504256K} only tested two basic models for the seed population. They did not vary the cut-off energy or the shape of the cut-off. For example, if the source of the plasma are radio galaxies, the break frequency of the population depends on the age of the plasma. A heavily aged seed population would have a strong, exponential cut-off spectrum which would translate into a curved integrated spectrum. However, as noted by \citet{2015arXiv150504256K}, the break in the spectrum is difficult to explain with synchrotron emission, as only a sharp break at $\gamma\sim10^{4-5}$ in the seed electron population can explain the spectrum. A scenario involving re-acceleration, namely one which preserves the break in the spectrum, is consistent with the observed spectrum. 

\subsubsection{Best scenarios?}

All in all, scenarios with strong downstream magnetic fields or involving  an aged electron population represent promising avenues for explaining the steepening of the spectrum. However more data and simulations are needed to explore these options. 

High-resolution images at $4$ GHz from VLA, which better resolve the `Sausage' relic, point out possible anisotropies in the brightness distribution not visible at current resolutions (e.g. filaments; van Weeren et al. in prep.). Such filaments could be related to variations in the magnetic field strength, while polarisation measurements from the same data will be able to reveal the magnetic structure in more detail. Simulations probing the full parameters space of cut-off energy of the seed electrons and the strength of the cut-off are required, in order to match observations. 

\section{Conclusions}\label{sec:conclusion}
In this paper, we have presented high frequency ($>2.5$ GHz) radio observations of the `Sausage' and `Toothbrush' relics. We combined these new measurements with GMRT and WSRT to study the relic spectrum over $2.5$ dex in frequency.

\begin{itemize}
\item  We detect the `Sausage' and `Toothbrush' relics at high-radio frequencies up to $30$ GHz. 
\item Using CARMA and AMI, we find compelling evidence for steepening in the high frequency spectrum of both relics. The low frequency ($<2.5$ GHz) spectrum of both the `Sausage' and `Toothbrush' relics is well described by a single power law with spectral index $\sim-0.9$. The high frequency spectrum ($>2.0$ GHz) is steeper than $\sim-1.45$ in both cases. The result hold when using predicted, total power flux densities. 
\item The \citet{1998A&A...332..395E} model cannot explain the observed spectrum. A possible explanation would be that the relics are formed through shock acceleration of seed relativistic electrons with sharp spectral breaks at $\sim$ Gev energies, e.g. pre-accelerated from past AGN activity.
\end{itemize}
The models currently tested are all DSA-based with variations in the seed population. However, the mismatch between the models and the observations suggests it is necessary we revisit the theory of radio relics. Therefore, the development of new theoretical models, building upon the observations presented here, is necessary. To attain this goal, high frequency data is crucial. To further remove uncertainties more measurements must be done and simpler, more compact relics are the ideal candidates to alleviate difficulties and minimise bias that may occur by missing flux when observing at high frequency. CARMA has been decommissioned. The spectral width of the observations presented here could be surpassed within the next few years when very low and very high frequency instruments are rolled out with accurate flux scales, high resolution (e.g. Low Frequency Array, LOFAR, international baselines) and large field of view (e.g. low bands of Atacama Large Millimeter/submillimeter Array, ALMA). Till then, instruments such as Effelsberg and the Green Bank Telescope will be able to give us CARMA-like resolution at $\sim30$ GHz.

\section*{Acknowledgements}
We would like to thank the anonymous referee for her/his useful comments which helped improve the clarity of the paper. We thank the staff of the Mullard Radio Astronomy Observatory for their invaluable assistance in the operation of AMI, which is supported by Cambridge University. Support for CARMA construction was derived from the states of California, Illinois, and Maryland, the James S. McDonnell Foundation, the Gordon and Betty Moore Foundation, the Kenneth T. and Eileen L. Norris Foundation, the University of Chicago, the Associates of the California Institute of Technology, and the National Science Foundation. Based in part on observations with the 100-m telescope of the MPIfR (Max-Planck-Institut f{\" u}r Radioastronomie) at Effelsberg. The Westerbork Synthesis Radio Telescope is operated by the ASTRON (Netherlands Institute for Radio Astronomy) with support from the Netherlands Foundation for Scientific Research (NWO). We thank the staff of the GMRT who have made these observations possible. GMRT is run by the National Centre for Radio Astrophysics of the Tata Institute of Fundamental Research. The National Radio Astronomy Observatory is a facility of the National Science Foundation operated under cooperative agreement by Associated Universities, Inc. This research has made use of NASA's Astrophysics Data System. AS and HR acknowledge financial support from NWO (grant number: NWO-TOP LOFAR 614.001.006). TS and HR acknowledges support from the ERC Advanced Investigator program NewClusters 321271. CR acknowledges the support of STFC studentships. RJvW is supported by NASA through the Einstein Postdoctoral grant number PF2-130104 awarded by the Chandra X-ray Center, which is operated by the Smithsonian Astrophysical Observatory for NASA under contract NAS8-03060. TWJ acknowledges support from NSF (USA) grant AST1211595. MH acknowledges support by the research group FOR 1254 founded by the Deutsche Forschungsgemeinschaft. JJH wishes to thank the Netherlands Institute for Radio Astronomy (ASTRON) for a postdoctoral fellowship.

\bibliographystyle{mn2e.bst}
\bibliography{Ultimate_spectrum}

\appendix
\section{Areas used for measuring integrated flux densities for the interferometric spectrum}

\begin{figure}
\begin{center}
\includegraphics[trim=0cm 0cm 0cm 0cm, width=0.495\textwidth]{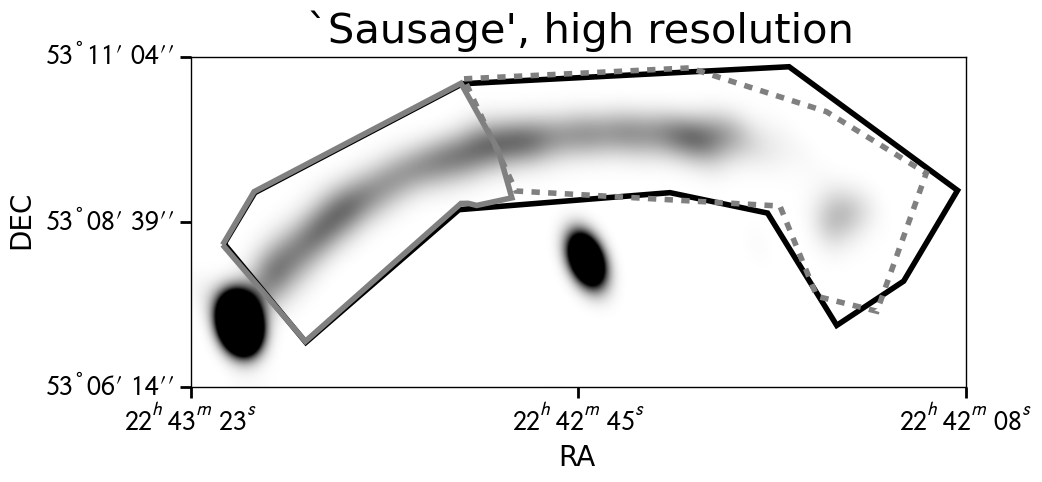}
\end{center}
\vspace{-10pt}
\caption{Area used for measuring the integrated spectrum of the `Sausage' radio relic. The area is large enough to capture diffuse emission in the downstream area of the shock, while avoiding contamination from radio galaxies (such as source H and B in Figure \ref{fig:mapS}).  The black line marks the area used for calculating the total spectrum. The solid gray line marks the area used for measuring the spectrum of the left side of the relic, while the dashed gray line marks the area use for the right side of the relic. These correspond to the spectra presented in Figure~\ref{fig:intspecS}.}
\label{fig:areaS}
\vspace{-10pt}
\end{figure}

\begin{figure}
\begin{center}
\includegraphics[trim=0cm 0cm 0cm 0cm, width=0.495\textwidth]{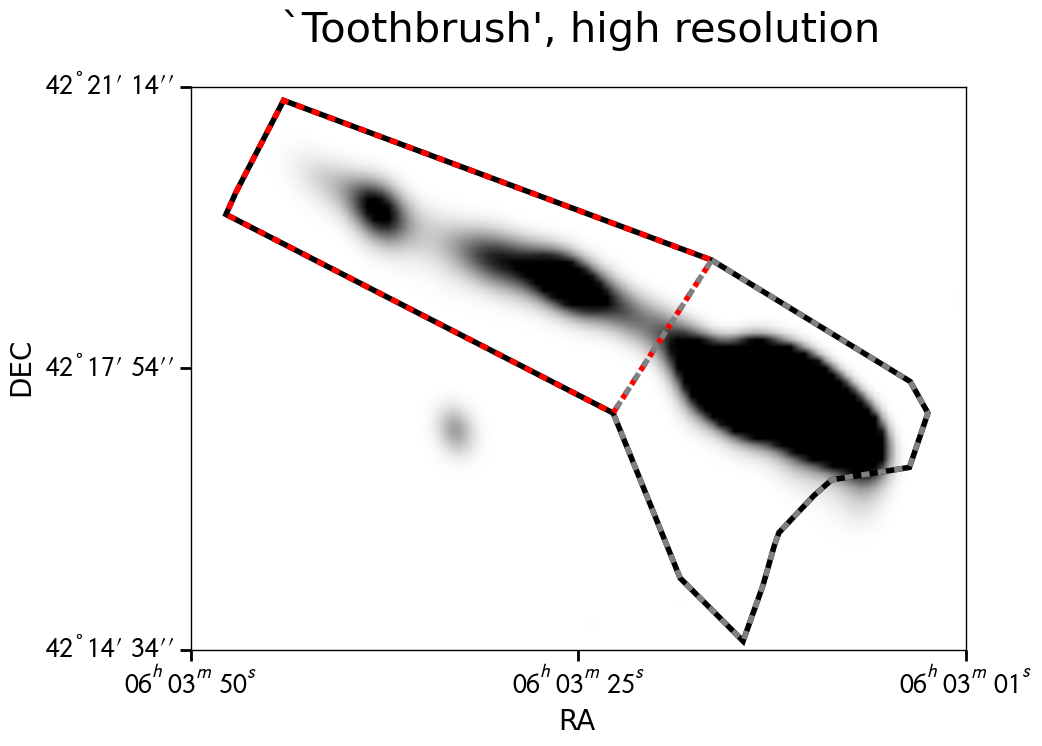}
\end{center}
\vspace{-10pt}
\caption{Area used for measuring the integrated spectrum of the `Toothbrush' radio relic. The area was designed to avoid a point source close to component B1, which is very bright at high frequencies (see Figure~\ref{fig:mapT}). The black line marks the area used for calculating the total spectrum (B1+B2+B3). The dashed red line marks the area used for measuring the spectrum of the left side of the relic (B2+B3), while the dashed gray line marks the area use for the right side of the relic (B1). These correspond to the spectra presented in Figure~\ref{fig:intspecT}.}
\label{fig:areaT}
\vspace{-10pt}
\end{figure}

In the comparable interferometric images with 800 $\lambda$ uv cut and uniform weighting, we use the same region at all frequencies for measuring the integrated spectra of the `Sausage' and `Toothbrush' relics, shown in Figures~\ref{fig:areaS} and \ref{fig:areaT}. The resulting flux density values are given in Tables \ref{tab:intspec} and ~\ref{tab:intspecsub}. Using the same region ensures the summing up of emission from electrons with all ranges of ages, from recently accelerated particles to particles injected $>100$ Myr in the past. This results in a spectrum measured given the assumptions laid out in \citet{1998A&A...332..395E}. Using too large an area would result in contamination from point sources or diffuse giant halo emission. However, too small an area would not fully probe the downstream area. 

\begin{table*}
\begin{center}
\caption{Radio spectrum of subsections of the `Sausage' and `Toothbrush' relics, using the regions described in Figure \ref{fig:areaS} and \ref{fig:areaT}.}
\vspace{-10pt}
\begin{tabular}{l r r r r r  r r r r r r r r r  }
\hline
\hline
Freq (GHz) & 0.15 & 0.24 & 0.325 & 0.61 & 1.2 & 1.4 & 1.7 & 2.25 & 2.3 & 2.75 & 3.25 & 3.75 & 16 & 30 \\  \hline
\multicolumn{5}{|l|}{`Sausage' left} \\ 
Flux density (mJy) & 293.7 & & 158.0 & 107.6 & 59.8 & 55.9 & 41.5 & 29.3 & 24.4 & 23.8 & 16.5 & 14.4 &  0.7 &  0.6 \\
Error (mJy) & 31.1 & & 16.4  & 10.8 & 6.0 &  5.6 & 4.2 & 1.8 & 2.5 &  1.3 & 0.9 & 1.4 & 0.3 & 0.2 \\
\hline
\multicolumn{5}{|l|}{`Sausage' right} \\ 
Flux density (mJy) & 482.2 & & 159.0 & 115.8 & 65.9 & 61.2 & 49.8 & 31.9 & 30.3 & 26.4 & 19.6 & 13.0 &  0.6 &  0.3\\
Error (mJy) & 50.0 & & 16.8 & 11.7 & 6.6 & 6.2 & 5.0 &  2.0 & 3.2 & 1.5 & 1.0 &  1.4 & 0.4 & 0.2 \\
\hline
\multicolumn{5}{|l|}{`Toothbrush' B1} \\ 
Flux density (mJy) & 2429.9 & 1101.2 &  795.7 & 554.1 & 251.8 & 216.8 & 173.0 &  & 116.1 & &  & &  7.6 &   3.0 \\
Error (mJy) & 243.1 & 110.2 & 79.6 & 55.4 & 25.2 & 21.7 & 17.3 & & 11.6 & &  & &  0.5 &  0.4  \\
\hline
\multicolumn{5}{|l|}{`Toothbrush' B2+B3} \\ 
Flux density (mJy) & 718.0 & 365.5 &  246.9 & 189. &  93.0 &  78.7 & 63.4 & & 46.7 & & & & 3.1 &  0.0 \\
Error (mJy) & 72.0 & 36.9 & 24.7 & 18.9 & 9.3 & 7.9 & 6.4 & & 4.7 & & & & 0.3 & 0.3\\
\hline
\end{tabular}
\label{tab:intspecsub}
\vspace{-10pt}
\end{center}
\end{table*}

\label{lastpage}
\end{document}